\documentclass[12pt,aps,prd,preprint,groupedaddress,tightenlines,nofootinbib,floatfix]{revtex4}

\usepackage[dvipdfm]{graphicx}

\newcommand{\bra}{\left\langle}
\newcommand{\ket}{\right\rangle}
\newcommand{\msbar}{\overline{\rm MS}}

\begin{document}

\begin{flushright}
{\normalsize UTHEP-625}\\
{\normalsize UTCCS-P-62}\\
\end{flushright}

\title{Charm quark system at the physical point of 2+1 flavor lattice QCD}

\author{Y.~Namekawa$^1$, S.~Aoki$^{1,2}$, K.~-I.~Ishikawa$^3$, N.~Ishizuka$^{1,2}$,
T.~Izubuchi$^4$, K.~Kanaya$^2$, Y.~Kuramashi$^{1,2,5}$,
M.~Okawa$^3$, Y.~Taniguchi$^{1,2}$,
A.~Ukawa$^{1,2}$, N.~Ukita$^1$ and T.~Yoshi\'e$^{1,2}$\\
(PACS-CS Collaboration)
}

\affiliation{
$^1$~Center for Computational Sciences, University of Tsukuba,
Tsukuba, Ibaraki 305-8577, Japan\\
$^2$~Graduate School of Pure and Applied Sciences, University of Tsukuba,
Tsukuba, Ibaraki 305-8571, Japan\\
$^3$~Graduate School of Science, Hiroshima University,
Higashi-Hiroshima, Hiroshima 739-8526, Japan\\
$^4$~Riken BNL Research Center, Brookhaven National Laboratory,
Upton, New York 11973, USA\\
$^5$~RIKEN Advanced Institute for Computational Science,
Kobe, Hyogo 650-0047, Japan
}

\date{\today}

\begin{abstract}
We investigate the charm quark system
using the relativistic heavy quark action 
on 2+1 flavor PACS-CS configurations
previously generated on
$32^3 \times 64$ lattice.  
The dynamical up-down and strange quark masses
are set to the physical values 
by using the technique of reweighting
to shift the quark hopping parameters
from the values employed in the configuration generation.
At the physical point, the lattice spacing equals 
$a^{-1}=2.194(10)$~GeV and the spatial extent $L=2.88(1)$ fm. 
The charm quark mass is determined by
the spin-averaged mass of the 1S charmonium state,
from which we obtain
$m_{\rm charm}^{\msbar}(\mu = m_{\rm charm}^{\msbar}) = 1.260(1)(6)(35)$~GeV,
where the errors are due to our statistics,
scale determination and renormalization factor.
An additional systematic error from the heavy quark is
of order $\alpha_s^2 f(m_Q a)(a \Lambda_{QCD})$,
which is estimated to be a percent level
if the factor $f(m_Q a)$ analytic in $m_Q a$ is of order unity. 
Our results for the charmed and charmed-strange meson
decay constants are 
$f_D=226(6)(1)(5)$~MeV, $f_{D_s}=257(2)(1)(5)$~MeV,
again up to the heavy quark errors
of order $\alpha_s^2 f(m_Q a)(a \Lambda_{QCD})$.
Combined with the CLEO values
for the leptonic decay widths,
these values yield 
$|V_{cd}| = 0.205(6)(1)(5)(9)$,
$|V_{cs}| = 1.00(1)(1)(3)(3)$,
where the last error is on account of
the experimental uncertainty of the decay widths.
\end{abstract}

\maketitle

\setcounter{equation}{0}
\section{Introduction}
\label{section:introduction}

Precise determination of the Cabibbo-Kobayashi-Maskawa (CKM)
quark mixing matrix is an indispensable step to establish
the validity range of the standard model,
and to search for new physics at higher energy scales.
Lattice QCD has been making steady progress in this direction.
For the matrix elements such as $|V_{ud}|$ and $|V_{us}|$
in the first row which involve only light quarks, 
dynamical simulations including up, down and strange quarks
have reached the point
where the relevant pseudoscalar meson decay constants
and form factors are being determined at subpercent precision.  
On the other hand, for $|V_{cd}|$ and $|V_{cs}|$ in the second row,
the precision of lattice QCD calculation for the decay constants
and form factors is still at 5 to 10\% level.
This is not clearly superior to non-lattice QCD determinations.
Indeed, the estimate quoted in Particle Data Group (PDG) 2010,
$|V_{cd}| = 0.230(11)$~\cite{PDG_2010} with an accuracy of 5\%,  
is obtained from neutrino and anti-neutrino experiments.
\footnote{
 $|V_{cs}|$ is hard to be estimated
 from neutrino and anti-neutrino experiments,
 $|V_{cs}| = 0.94^{+0.32}_{-0.26} \pm 0.13$~\cite{PDG_2010}.
}
Much effort is needed on the part of lattice QCD
toward a better precision in the charm sector.

One of the difficulties with the charm quark in lattice QCD simulations
at a typical cutoff $a^{-1} \approx 2$~GeV 
resides in significant cutoff errors due to the charm quark mass.
The heavy quark mass correction is $m_Q a \sim 1$,
and hence we must control errors to all orders of $m_Q a$
to achieve a few percent accuracy.
The Fermilab action~\cite{Fermilab_action} and
the relativistic heavy quark action~\cite{RHQ_action_Tsukuba,RHQ_action_Columbia}
have been proposed to meet this goal.
In the present work we employ the relativistic formalism of
Ref.~\cite{RHQ_action_Tsukuba} to explore the charm quark system.

Another source that prevents precise evaluations in lattice QCD 
is the error associated with chiral extrapolations
in the light quark masses.
This problem has been increasingly alleviated through progress
toward simulations with lighter and lighter dynamical quark masses
and sophisticated application of chiral perturbation theory techniques. 
The acceleration of dynamical lattice QCD simulation
using multi-time steps for infrared and ultraviolet
modes~\cite{DDHMC,mass_precondition} has made it possible
to run simulations with light up, down and strange quark masses
around their physical values~\cite{PACS_CS_0}.   
In such simulations, uncertainties due to chiral extrapolations
are drastically reduced.

In fact we can proceed one more step and reweight~\cite{reweight} dynamical simulations
such that dynamical quark masses take exactly 
the physical values.
A potential difficulty with dynamical lattice QCD 
is a large fluctuation of quark determinant ratios necessary
for reweighting.
We have demonstrated the feasibility of this procedure
in Ref.~\cite{PACS_CS_1} by reweighting a set of PACS-CS configurations 
with $m_\pi=152(6)$~MeV and $m_K=509(2)$~MeV
to those
with $m_\pi=135(6)$~MeV and $m_K=498(2)$~MeV.
Once the reweighting is successfully made,
ambiguities associated with chiral extrapolations are
completely removed.
In the present work we employ the reweighting factors
and the set of original dynamical configurations 
employed in Ref.~\cite{PACS_CS_1}.
Hence our light quark masses sit at the physical point.

In this paper we present our work for the charm quark system
treated with the relativistic heavy quark formalism~\cite{RHQ_action_Tsukuba} 
on the 2+1 dynamical flavor PACS-CS configurations
of $32^3 \times 64$ lattice generated 
with the Wilson-clover quark
and reweighted to the physical point for 
up, down and strange quark masses. 
The lattice spacing is estimated as $a^{-1}=2.194(10)$~GeV.
We measure the masses and decay constants of charmonia,
charmed mesons and charmed-strange mesons.
We then calculate the charm quark mass and the CKM matrix elements.

This paper is organized as follows.
Section II explains our method and simulation parameters.
Section III describes our results
for the charmonium spectrum and the charm quark mass.
In Sec. IV, we show our charmed meson and charmed-strange meson spectrum.
Section V is devoted to present
our pseudoscalar decay constants
and the CKM matrix elements.
Our conclusions are given in Sec. VI.

\setcounter{equation}{0}
\section{Set up}
\label{section:setup}

Our calculation is based on a set of $N_f=2+1$ flavor
dynamical lattice QCD configurations
generated by the PACS-CS Collaboration~\cite{PACS_CS_1}
on a $32^3\times 64$ lattice 
using the nonperturbatively $O(a)$-improved Wilson quark action 
with $c_{\rm SW}^{\rm NP}=1.715$~\cite{Csw_NP}
and the Iwasaki gauge action~\cite{RG} at $\beta=1.90$.
The aggregate of 2000 MD time units were generated at the hopping parameter
given by $(\kappa_{ud}^0,\kappa_{s}^0)=(0.13778500, 0.13660000)$,
and 80 configurations at every 25 MD time units were used for measurements.
We then reweight those configurations to the physical point
given by $(\kappa_{ud},\kappa_{s})=(0.13779625, 0.13663375)$.
The reweighting shifts the masses of $\pi$ and $K$ mesons 
from $m_\pi=152(6)$~MeV and $m_K=509(2)$~MeV
  to $m_\pi=135(6)$~MeV and $m_K=498(2)$~MeV,
with the cutoff at the physical point estimated
to be $a^{-1}=2.194(10)$~GeV.

Observables at the physical point are evaluated
through the formula 
\begin{eqnarray}
 \bra {\cal O}[U](\kappa_{\rm ud},\kappa_{\rm s}) 
 \ket_{(\kappa_{\rm ud},\kappa_{\rm s})} 
 =
 \frac{\bra {\cal O}[U](\kappa_{\rm ud},\kappa_{\rm s})
            R_{\rm ud}[U] R_{\rm s}[U]
       \ket_{(\kappa_{\rm ud}^0,\kappa_{\rm s}^0)} }
      {\bra R_{\rm ud}[U] R_{\rm s}[U]
       \ket_{(\kappa_{\rm ud}^0,\kappa_{\rm s}^0)} },
 \label{eq:reweight}
\end{eqnarray}
where the reweighting factors are defined as
\begin{eqnarray}
 R_{\rm ud}[U]
 &=& \left\vert \det \left[ \frac{D_{\kappa_{\rm ud}}[U]}
                                 {D_{\kappa_{\rm ud}^0}[U]}
                     \right]
     \right\vert^2,\\
 R_{\rm  s}[U]
 &=& \det \left[ \frac{D_{\kappa_{\rm s}}[U]}
                      {D_{\kappa_{\rm s}^0[U]}}
          \right],
\end{eqnarray}
and $D_{\kappa_q}[U]$ is the Wilson-clover quark operator
with the hopping parameter $\kappa_q$.
We refer to Ref.~\cite{PACS_CS_1} for details of
our evaluation of the determinant ratio. 
Our parameters and statistics
at the physical point are collected in Table~\ref{table:statistics}.

The relativistic heavy quark formalism~\cite{RHQ_action_Tsukuba}
is designed to reduce cutoff errors of $O((m_Q a)^n)$
with arbitrary order $n$ to $O(f(m_Q a)(a \Lambda_{QCD})^2)$,
once all of the parameters in the relativistic heavy quark action
are determined nonperturbatively,
where $f(m_Q a)$ is an analytic function
around the massless point $m_Q a = 0$.
The action is given by
\begin{eqnarray}
 S_Q
 &=& \sum_{x,y}\overline{Q}_x D_{x,y} Q_y,\\
 D_{x,y}
 &=& \delta_{xy}
     - \kappa_{Q}
       \sum_i \left[  (r_s - \nu \gamma_i)U_{x,i} \delta_{x+\hat{i},y}
                     +(r_s + \nu \gamma_i)U_{x,i}^{\dag} \delta_{x,y+\hat{i}}
              \right]
     \nonumber \\
 &&  - \kappa_{Q}
              \left[  (r_t - \nu \gamma_i)U_{x,4} \delta_{x+\hat{4},y}
                     +(r_t + \nu \gamma_i)U_{x,4}^{\dag} \delta_{x,y+\hat{4}}
              \right]
     \nonumber \\
 &&  - \kappa_{Q}
              \left[   c_B \sum_{i,j} F_{ij}(x) \sigma_{ij}
                     + c_E \sum_i     F_{i4}(x) \sigma_{i4}
              \right],
\end{eqnarray}
where $\kappa_Q$ is the hopping parameter for the heavy quark.
The parameters $r_t, r_s, c_B, c_E$ and $\nu$ are adjusted as follows.
We are allowed to choose $r_t=1$,
and we employ a one-loop perturbative value for $r_s$~\cite{RHQ_parameters}.
For the clover coefficients $c_B$ and $c_E$,
we include the non-perturbative contribution
in the massless limit $c_{\rm SW}^{\rm NP}$
for three flavor dynamical QCD~\cite{Csw_NP},
and calculate the heavy quark mass dependent contribution 
to one-loop order in perturbation theory~\cite{RHQ_parameters}
according to
\begin{eqnarray}
 c_{B,E}=(c_{B,E}(m_Q a) - c_{B,E}(0))^{\rm PT} + c_{\rm SW}^{\rm NP}.
\end{eqnarray}
The parameter $\nu$ is determined non-perturbatively
to reproduce the relativistic dispersion relation for
the spin-averaged  $1S$ states of the charmonium.
Writing 
\begin{equation}
   E({\vec p})^2
 = E({\vec 0})^2+c_{\rm eff}^2 |{\vec p}|^2,
\end{equation}
for $|{\vec p}| = 0, (2 \pi / L), \sqrt{2} (2 \pi / L)$,
and demanding the effective speed of light $c_{\rm eff}$
to be unity, we find $\nu=1.1450511$ with which
we have $c_{\rm eff} = 1.002(4)$.
It is noted that the remaining cutoff errors are
$\alpha_s^2 f(m_Q a)(a \Lambda_{QCD})$,
instead of $f(m_Q a)(a \Lambda_{QCD})^2$,
due to the use of one-loop perturbative values in part
for the parameters of our heavy quark action.

We tune the heavy quark hopping parameter to reproduce  
an experimental value of the mass for the spin-averaged $1S$ states of the charmonium,
given by
\begin{eqnarray}
 M(1S)^{exp}
 = (M_{\eta_c} + 3 M_{J/\psi})/4
 = 3.0678(3) \mbox{~GeV~\cite{PDG_2010}}.
\end{eqnarray}
This leads to $\kappa_{\rm charm}=0.10959947$
for which our lattice QCD measurement
yields the value $M(1S)^{lat} = 3.067(1)(14)$~GeV,
where the first error is statistical,
and the second is a systematic
from the scale determination.
Our parameters for the relativistic heavy quark action
are summarized in Table~\ref{table:input_parameters_for_RHQ}.

We use the following standard operators to obtain meson masses,
\begin{eqnarray}
 M_{\Gamma}^{fg}(x) = \bar{q}_f(x) \Gamma q_g(x),
\end{eqnarray}
where $f,g$ are quark flavors and
$\Gamma = I, \gamma_5, \gamma_{\mu}, i \gamma_{\mu} \gamma_5,
 i[\gamma_{\mu},\gamma_{\nu}]/2$.
The meson correlators are calculated
with a point and exponentially smeared sources
and a local sink.
The smearing function is given by $\Psi(r) = A \exp(-B r)$ at $r \not = 0$ and
$\Psi(0)=1$.
We set
$A = 1.2$, $B = 0.07$ for the $ud$ quark,
$A = 1.2$, $B = 0.18$ for the strange quark, and 
$A = 1.2$, $B = 0.55$ for the charm quark.
The number of source points is quadrupled
and polarization states are averaged
to reduce statistical fluctuations.
Statistical errors are analyzed by the jackknife method
with a bin size of 100 MD time units (4 configurations),
as in the light quark sector~\cite{PACS_CS_1}.

\begin{figure}[t]
\begin{center}
 \includegraphics[width=75mm]{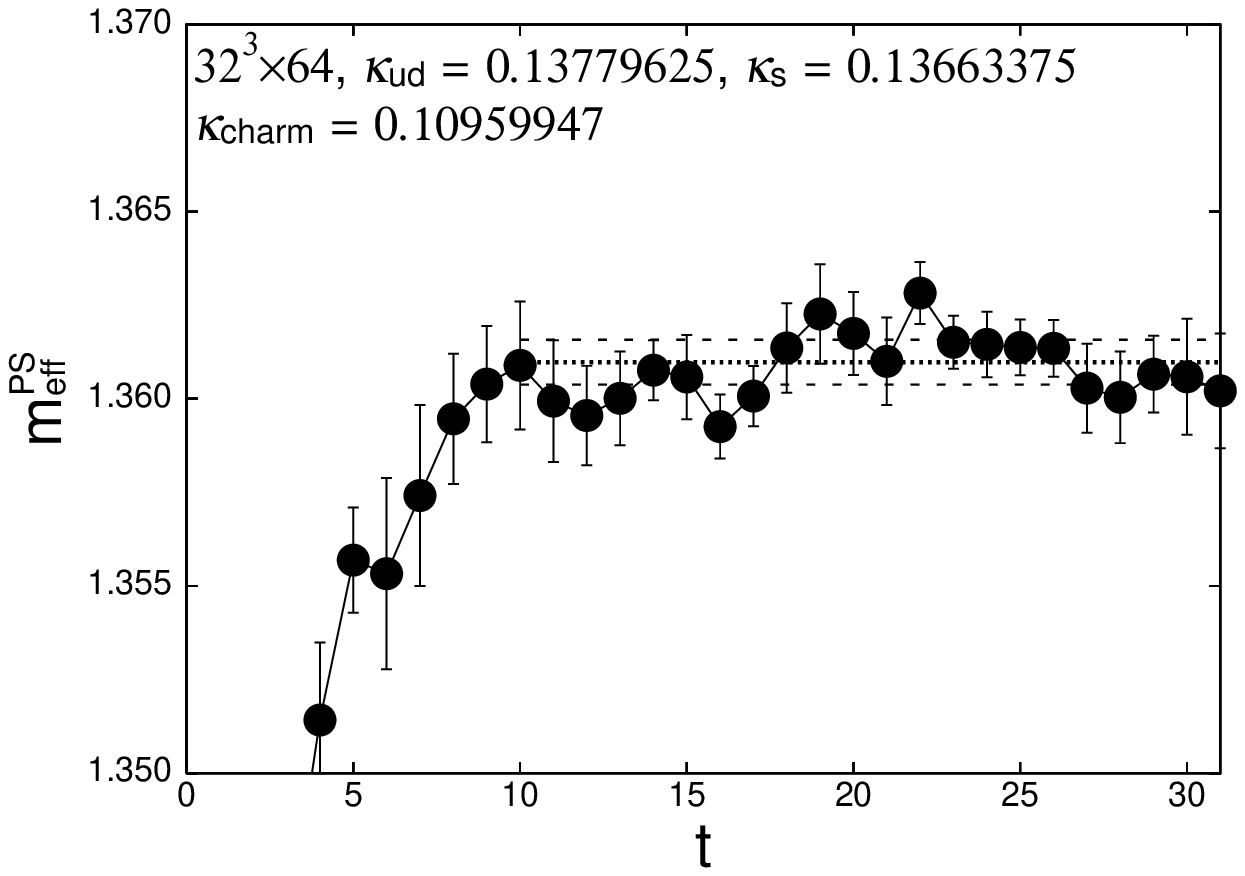}
 \includegraphics[width=75mm]{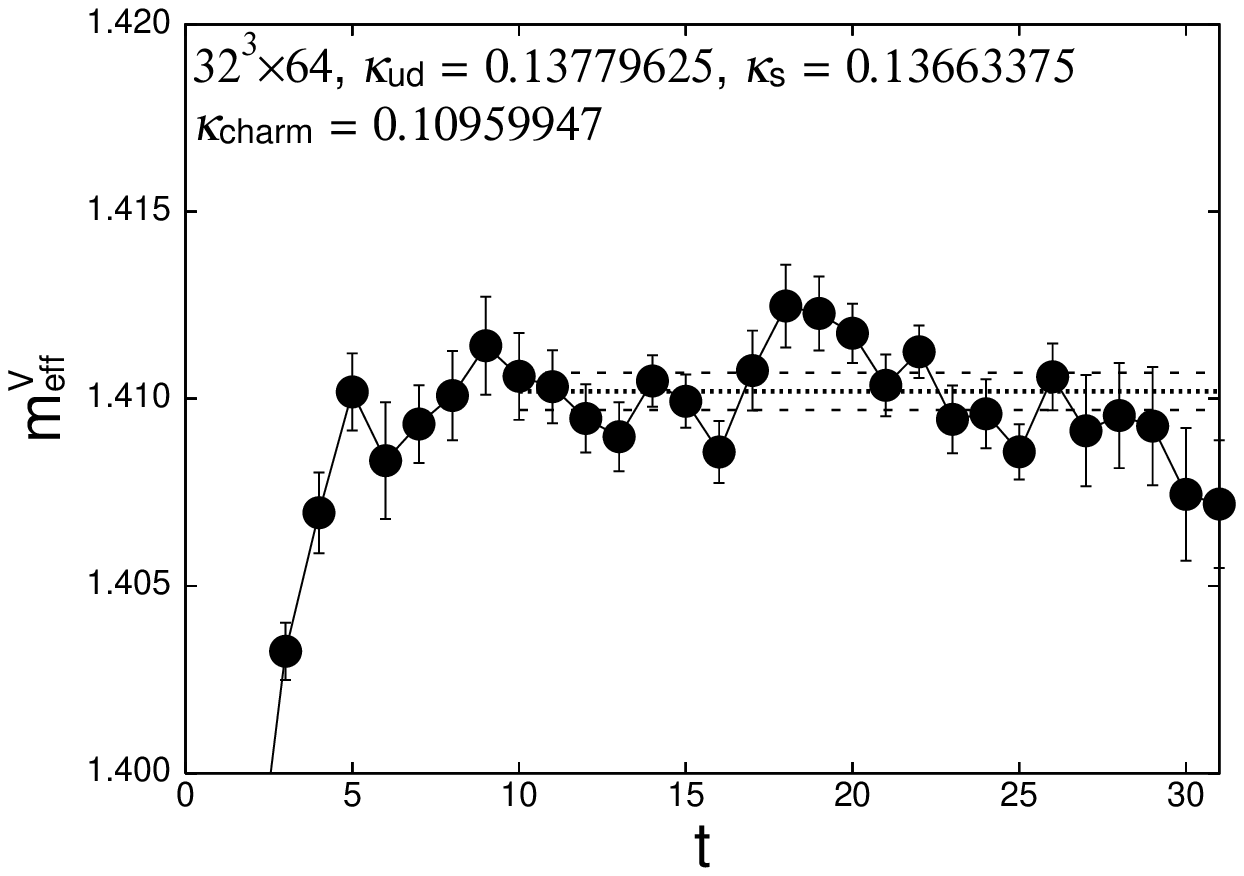}
 \includegraphics[width=75mm]{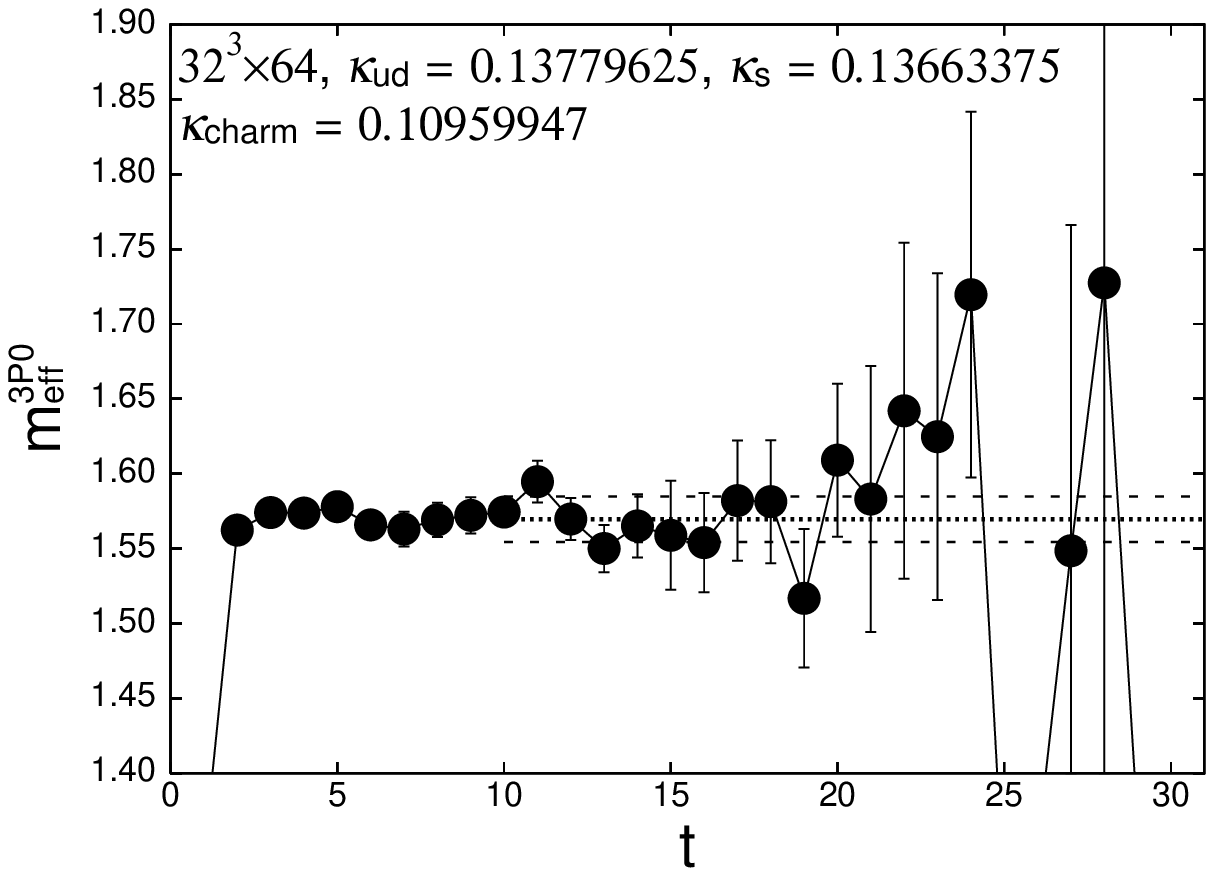}
 \includegraphics[width=75mm]{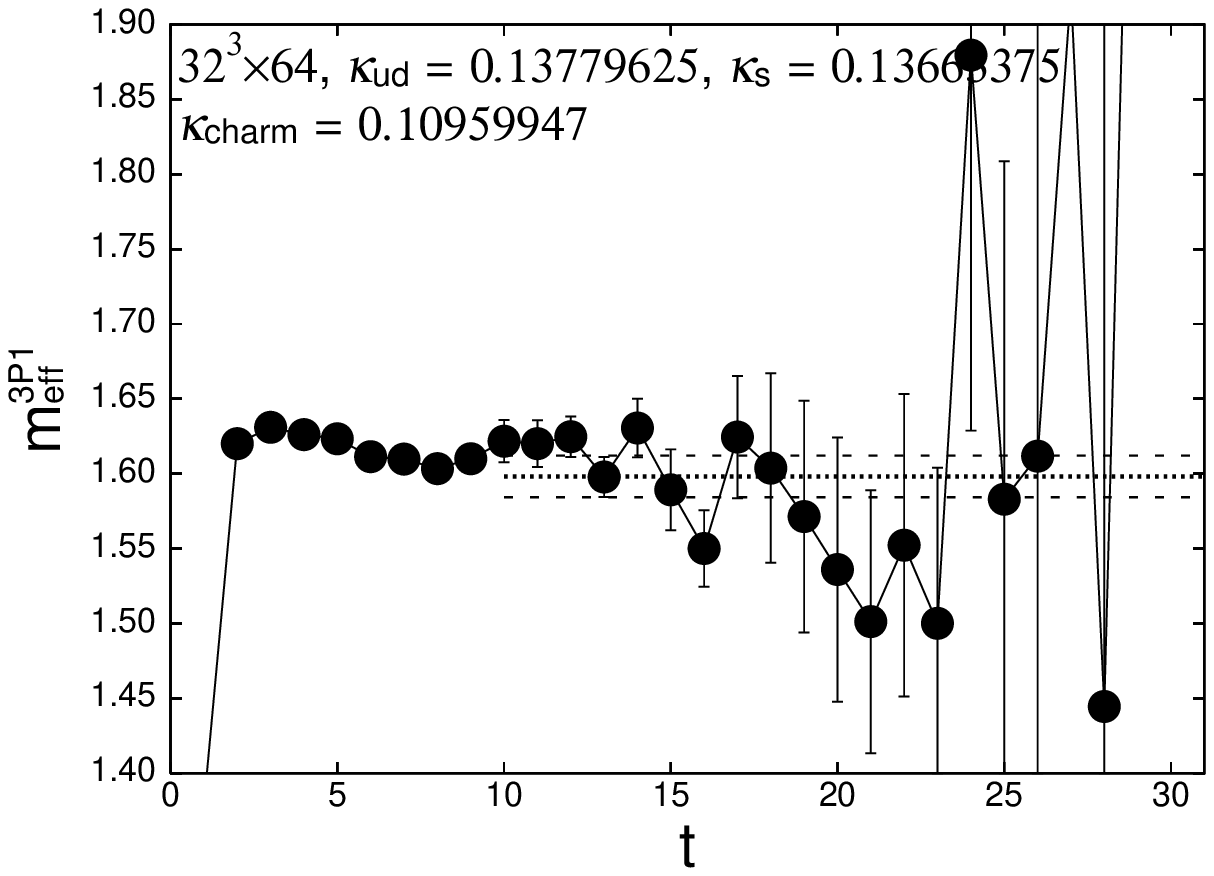}
 \includegraphics[width=75mm]{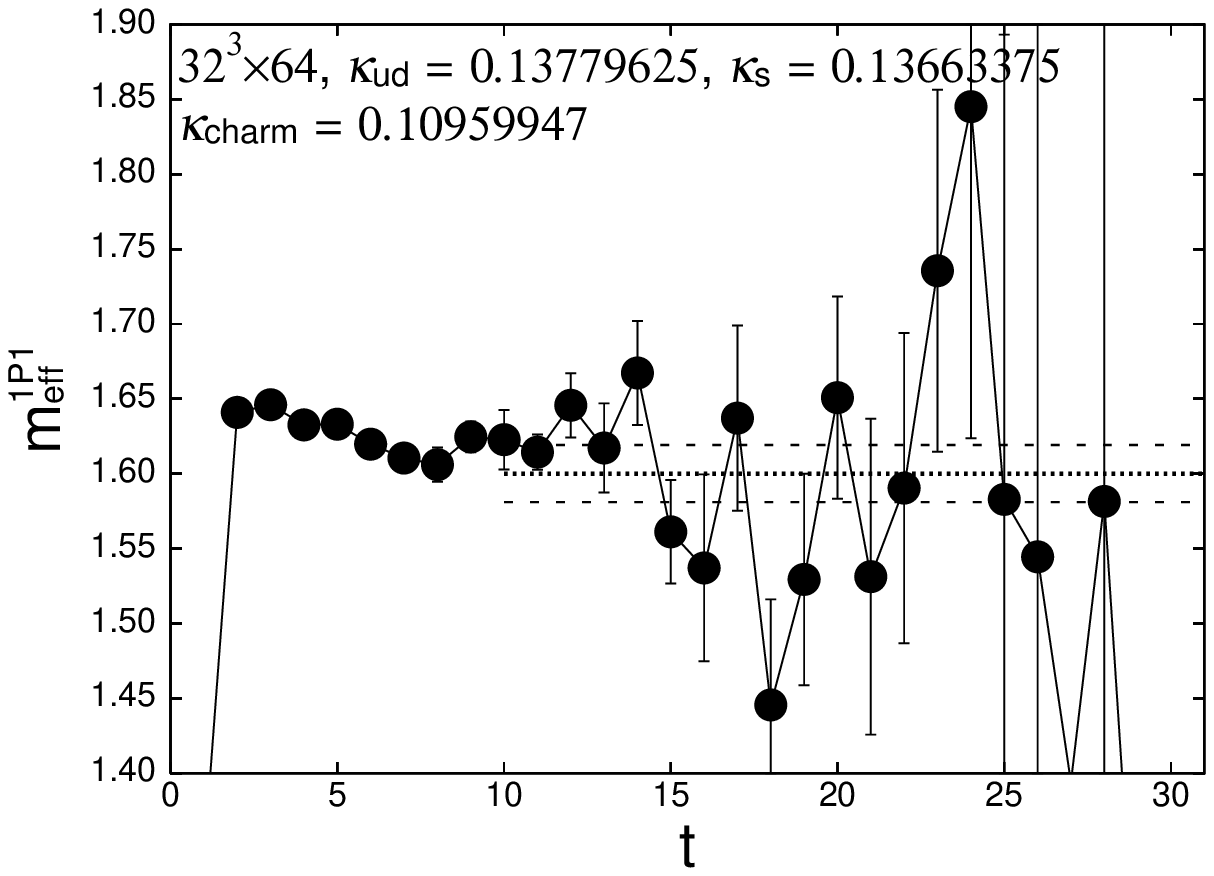}
 \caption{
 Effective masses for charmonium.
 }
 \label{figure:m_eff_charmonium}
\end{center}
\end{figure}

\begin{figure}[t]
\begin{center}
 \includegraphics[width=75mm]{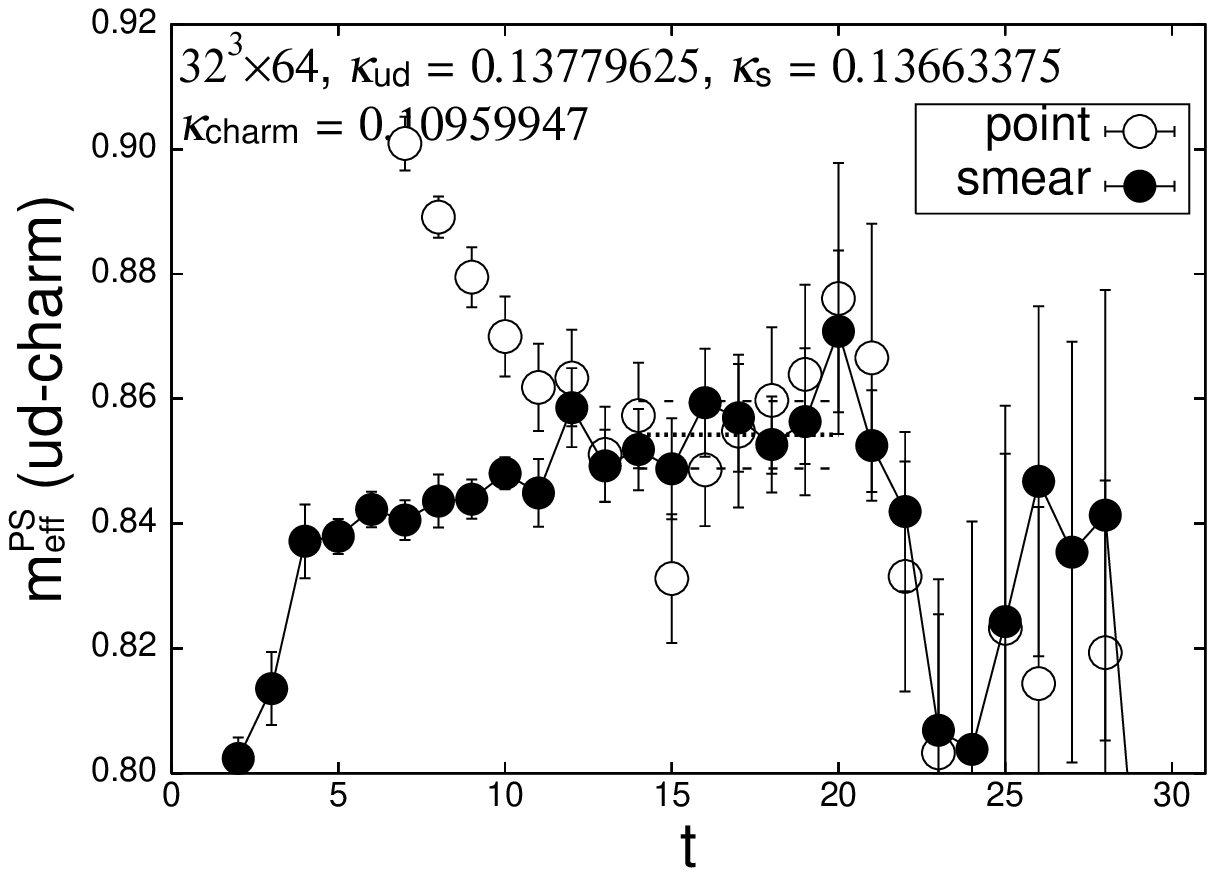}
 \includegraphics[width=75mm]{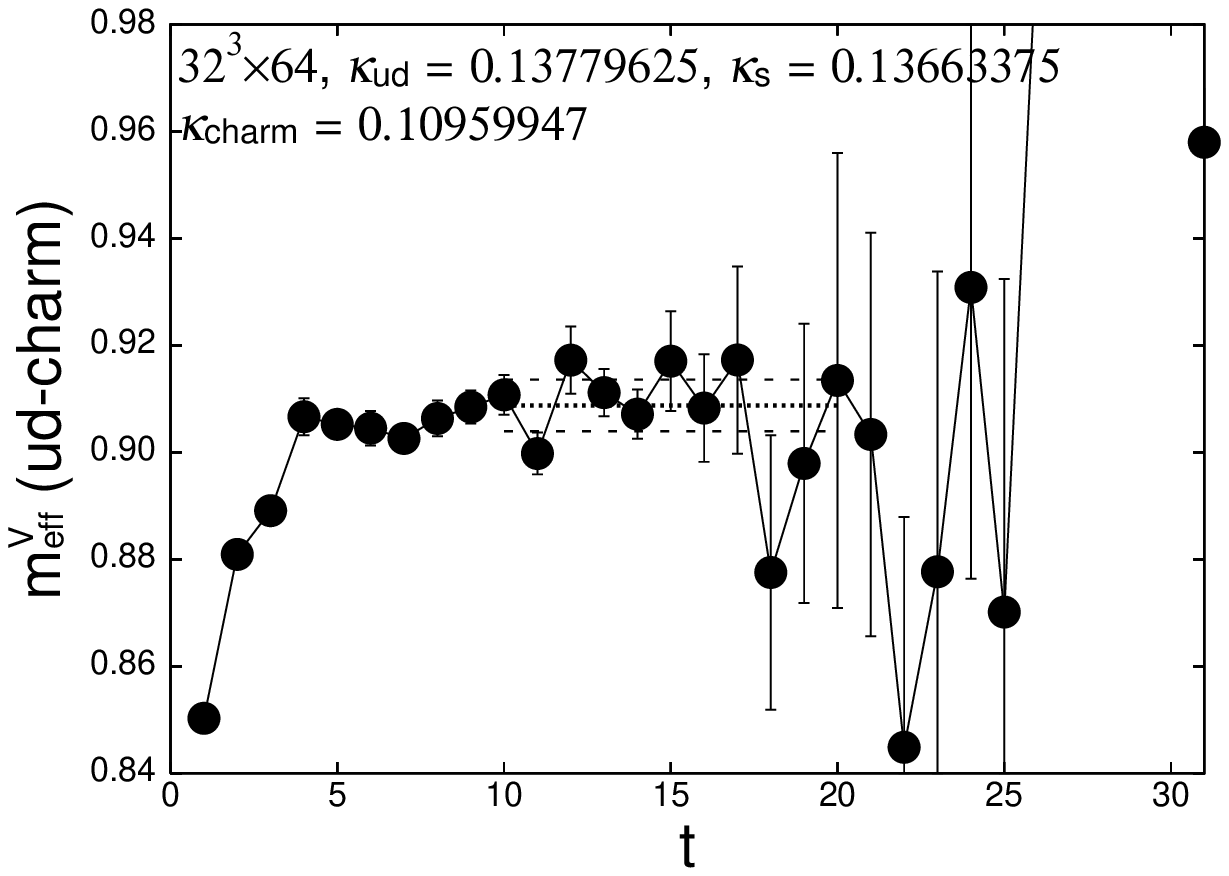}
 \caption{
 Effective masses for charmed mesons.
 }
 \label{figure:m_eff_ud_charm}
\end{center}
\end{figure}

\begin{figure}[t]
\begin{center}
 \includegraphics[width=75mm]{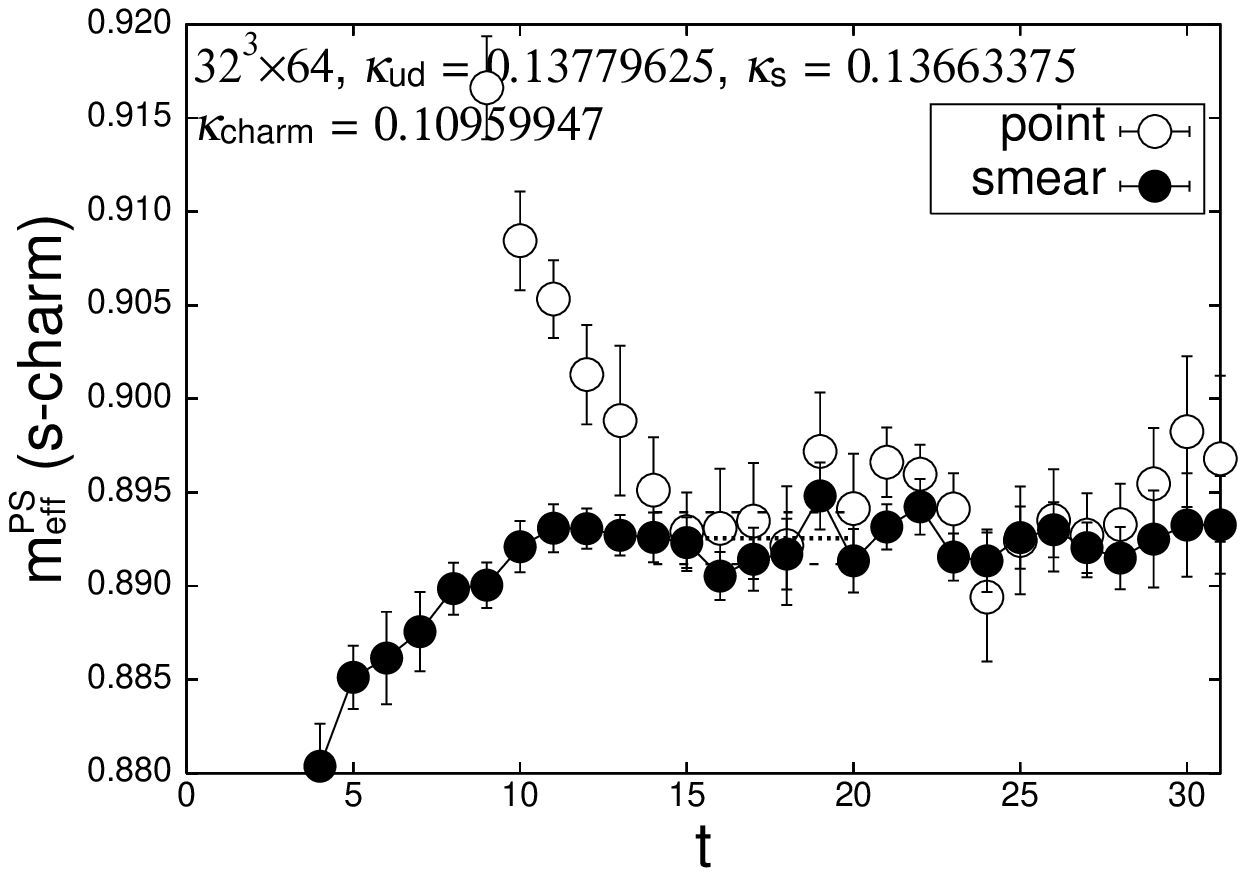}
 \includegraphics[width=75mm]{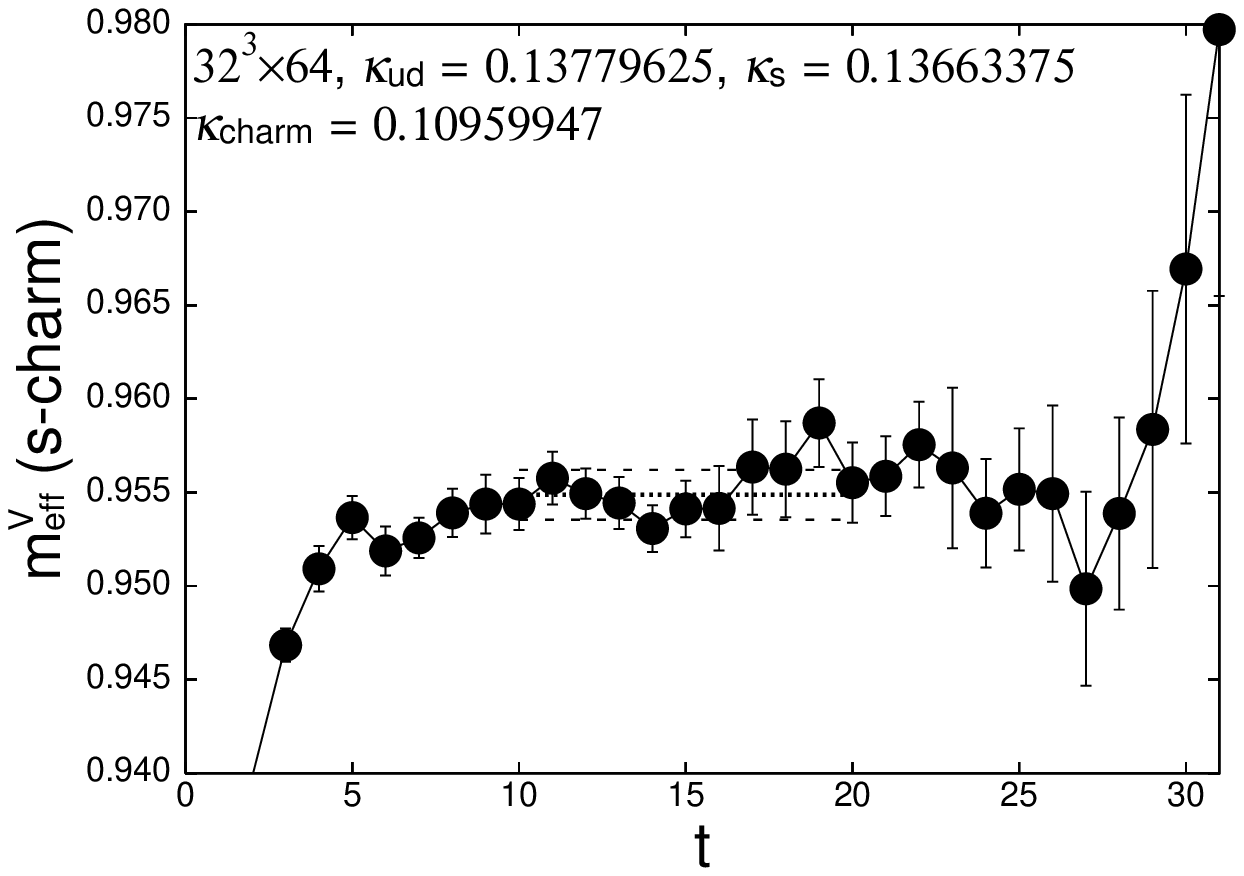}
 \includegraphics[width=75mm]{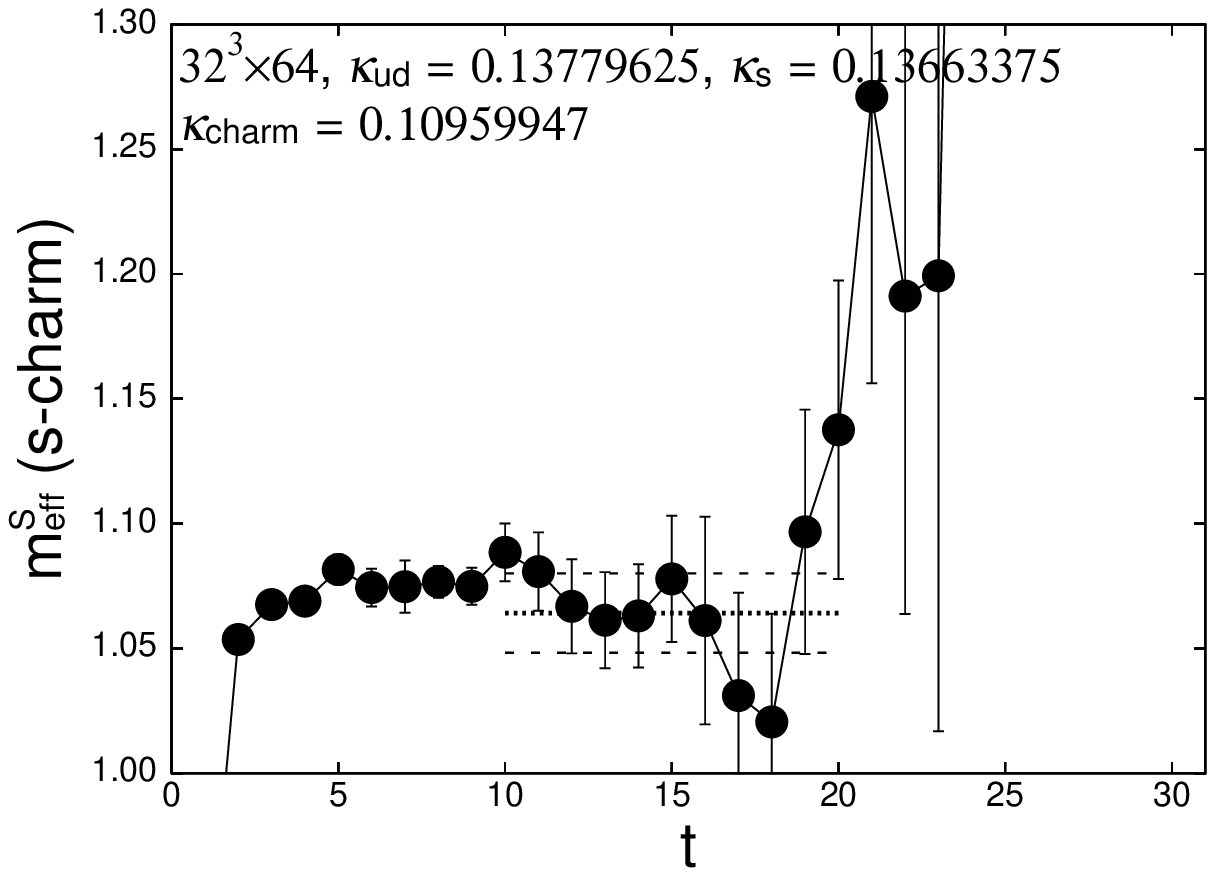}
 \includegraphics[width=75mm]{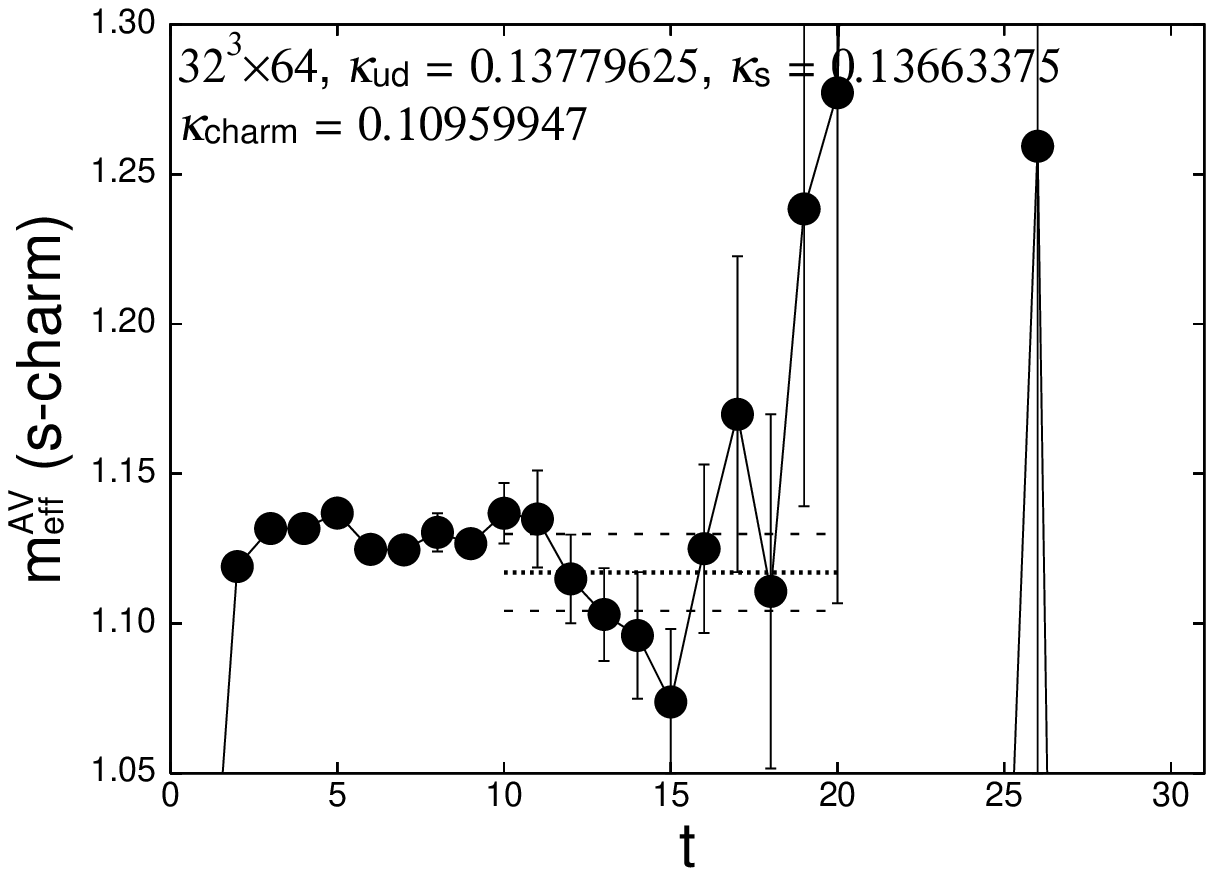}
 \caption{
 Effective masses for charmed-strange mesons.
 }
 \label{figure:m_eff_s_charm}
\end{center}
\end{figure}

We extract meson masses by fitting correlators
with a hyperbolic cosine function.
For charmonium, Fig.~\ref{figure:m_eff_charmonium}
shows effective masses,
from which we choose the fitting range to be
$[t_{min},t_{max}] = [10,32]$.
Similarly, Fig.~\ref{figure:m_eff_ud_charm}
and Fig.~\ref{figure:m_eff_s_charm}
represent effective masses for 
charmed mesons and charmed-strange mesons.
We employ the fitting range $[t_{min},t_{max}] = [14,20]$
for pseudoscalar mesons, and
$[t_{min},t_{max}] = [10,20]$ for the other channels.

We calculate the decay constant $f_{PS}$ of
the heavy-light pseudoscalar meson 
using the improved axial vector current $A_{4}^{imp}$.
\begin{eqnarray}
 i f_{PS} p_{\mu}
 &=& \langle 0 | A_{\mu}^{imp} | PS(p) \rangle,
 \label{equation:f_PS}
 \\
 A_{4}^{imp}
 &=& \sqrt{2 \kappa_q} \sqrt{2 \kappa_Q} Z_{A_{4}}
     \left\{ \bar{q}(x) \gamma_{4} \gamma_{5} Q(x)
     \right. \nonumber \\
 &&  \left. +c_{A_{4}}^{+}   \partial_{4}^{+}
             \left( \bar{q}(x) \gamma_{5} Q(x) \right)
            +c_{A_{4}}^{-}   \partial_{4}^{-}
             \left( \bar{q}(x) \gamma_{5} Q(x) \right)
     \right\},
\end{eqnarray}
where $\vert PS \rangle$ is the pseudoscalar meson state
and $\partial^{\pm}$ is the lattice forward and backward derivative.
For the renormalization factor $Z_{A_{4}}$
and the improvement coefficients of the axial current
$c_{A_{4}}^{+}$ and $c_{A_{4}}^{-}$,
we employ one-loop perturbation theory
to evaluate the mass-dependent contributions~\cite{Z_factors}, 
adding the nonperturbative contributions in the chiral limit by
\begin{eqnarray}
 c_{A_4}^{+} &=& (c_{A_4}^{+}(m_Q a) - c_{A_4}^{+}(0))^{\rm PT}
                 + c_{A}^{\rm NP}, \\
 Z_{A_4}     &=& (Z_{A_4}(m_Q a)     - Z_{A_4}(0))^{\rm PT} + Z_{A}^{\rm NP},
\end{eqnarray}
with $c_{A}^{\rm NP} = -0.03876106$~\cite{NP-c_A} and 
$Z_{A}^{\rm NP} = 0.781(20)$~\cite{PACS_CS_2}.

The bare quark mass is determined
through the axial vector Ward-Takahashi identity,
\begin{eqnarray}
 m_f^{AWI} + m_g^{AWI}
 = m_{PS}
   \frac{\bra 0 | A_4^{imp} | PS \ket}
        {\bra 0 | P         | PS \ket},
\end{eqnarray}
where $P$ is the pseudoscalar meson operator.
The renormalized quark mass in the $\msbar$ scheme
is given by
\begin{eqnarray}
 m_f^{\msbar}(\mu) = Z_m(\mu) m_f^{AWI}.
 %
 %
 \label{equation:m_quark}
\end{eqnarray}
Similar to the case of $Z_{A_4}$,
the renormalization factor for the quark mass
at the renormalization scale $\mu$, $Z_m (\mu)$,
is nonperturbatively determined at the massless point,
\begin{eqnarray}
 Z_{m}(\mu) = (Z_{m}(m_Q a) - Z_{m}(0))^{\rm PT}(\mu) + Z_{m}^{\rm NP}(\mu),
\end{eqnarray}
with $Z_{m}^{\rm NP}(\mu=1/a) = 1.308(35)$~\cite{PACS_CS_2}.
The charm quark mass is then evolved to $\mu = m_{\rm charm}^{\msbar}$
using $N_f=3$ four-loop beta function~\cite{beta_function}.
We employ $N_f=3$ based on the fact that
our simulation includes $N_f=2+1$ dynamical quarks.

\setcounter{equation}{0}
\section{Charmonium spectrum and charm quark mass}
\label{section:result_1}

\begin{figure}[t]
\begin{center}
 \includegraphics[width=75mm]{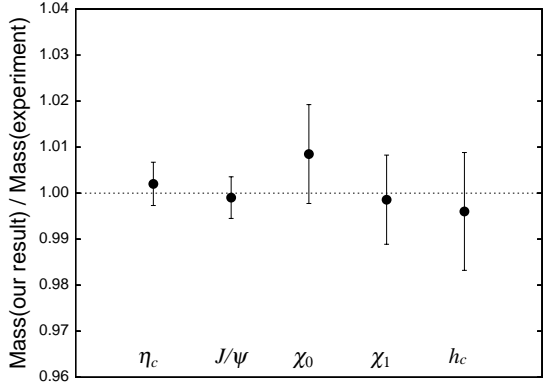}
 \caption{
 Our results for the charmonium mass spectrum
 normalized by the experimental values.
 }
 \label{figure:mass_charmonium_all}
\end{center}
\end{figure}

Our results for the charmonium spectrum
on the physical point are summarized in
Fig.~\ref{figure:mass_charmonium_all}
and Table~\ref{table:mass_charmonium}.
Within the error of 0.5--1\%,
the predicted spectrum is in reasonable agreement with experiment.

\begin{figure}[t]
\begin{center}
 \includegraphics[width=75mm]{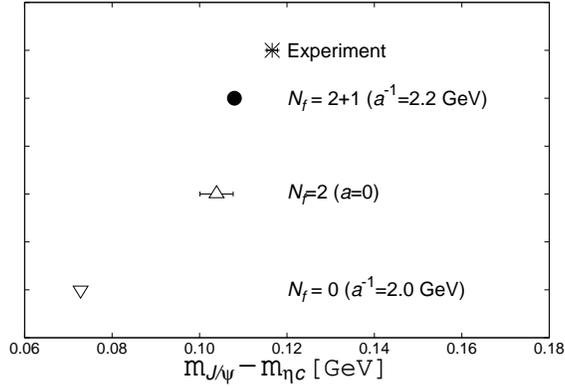}
 \caption{
 Hyperfine splitting of the charmonium
 with different number of flavors.
 }
 \label{figure:m_V_minus_m_PS}
\end{center}
\end{figure}

Let us consider the $1S$ states more closely.
Since these states are employed to tune the charm quark mass, 
the central issue here is the magnitude of the hyperfine splitting.
Our result $m_{J/\psi}-m_{\eta_c}=0.108(1)(0)$~GeV,
where the first error is statistical
and the second error is systematic from the scale determination,
is 7\% smaller than the experimental value of 0.117~GeV.  
In Fig.~\ref{figure:m_V_minus_m_PS},
we compare the present result
on $N_f=2+1$ flavor dynamical configurations 
with previous attempts on $N_f=2$ dynamical
and quenched configurations
using the same heavy quark formalism
and the Iwasaki gluon action~\cite{RHQ-N_f_0_2}.
We observe a clear trend that
incorporation of dynamical light quark effects 
improves the agreement.

We should note that the continuum extrapolation is to be performed.
A naive order counting implies that effects of
$O(\alpha_s^2 f(m_Q a)(a \Lambda_{QCD}))$
from the relativistic heavy quark action
is at a percent level.
Another aspect is that
dynamical charm quark effects
and disconnected loop contributions,
albeit reported to give a shift of only a few MeV~\cite{disconnected},
are not included in the present work.
Additional calculations are needed to draw a definite conclusion
for the hyperfine splitting of the charmonium spectrum.

\begin{figure}[t]
\begin{center}
 \includegraphics[width=75mm]{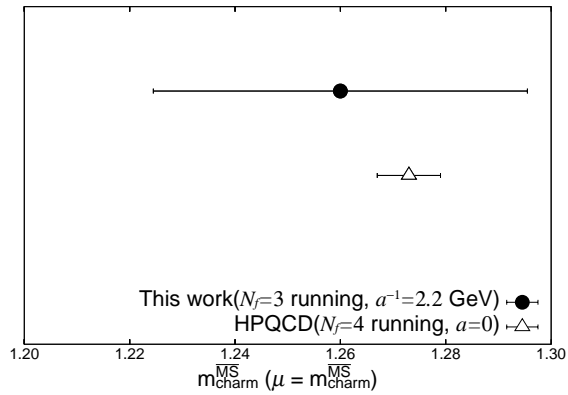}
 \caption{
 Comparison of the charm quark mass.
 The charm quark mass is obtained at $\mu = a^{-1}$,
 and evolved to $\mu = m_{charm}^{\msbar}$
 using four-loop beta function~\cite{beta_function}.
 We employ $N_f=3$ running based on the fact that
 our simulation includes $N_f=2+1$ dynamical quarks,
 while HPQCD collaboration uses $N_f=4$
 reflecting fictitious dynamical charm quark effects~\cite{HPQCD_1}.
 }
 \label{figure:m_charm}
\end{center}
\end{figure}

Using Eq.~(\ref{equation:m_quark}),
the charm quark mass is obtained as 
\begin{eqnarray}
 m_{\rm charm}^{\msbar}(\mu = m_{\rm charm}^{\msbar})  = 1.260(1)(6)(35) \mbox{~GeV},
\end{eqnarray}
where the first error is statistical,
the second is systematic from the scale determination,
and the third from uncertainty in the renormalization factor.  
The systematic error due to the heavy quark of
$O(\alpha_s^2 f(m_Q a)(a \Lambda_{QCD}))$ 
is also to be estimated.
Figure~\ref{figure:m_charm} compares our result
with a recent $N_f=2+1$ lattice QCD estimation
by the HPQCD Collaboration~\cite{HPQCD_1} 
in the continuum limit, 
which uses the HISQ form of the staggered quark action
for the heavy quark on the MILC dynamical configurations.

\setcounter{equation}{0}
\section{Charmed meson and charmed-strange meson spectrum}
\label{section:result_2}

We calculate the charmed meson
and charmed-strange meson masses
which are stable on our lattice with the spatial size of $L = 2.88(1)$~fm
and a lattice cutoff of $a^{-1}=2.194(10)$~GeV. 
The $D^*$ and $D_s^*$ meson decay channels are not open
in our lattice setup.
$D_{s0}^*$ and $D_{s1}$ meson masses are
below the $D K$ threshold~\cite{PDG_2010} but above 
the $D_s \pi$ threshold.
Their decays, however, 
are prohibited by the isospin symmetry.
On the other hand,
$D_{0}^*$ and $D_1$ meson masses are not computed
since their decay channels are open,
and therefore a calculation involving $D \pi$ contributions is needed.

\begin{figure}[t]
\begin{center}
 \includegraphics[width=75mm]{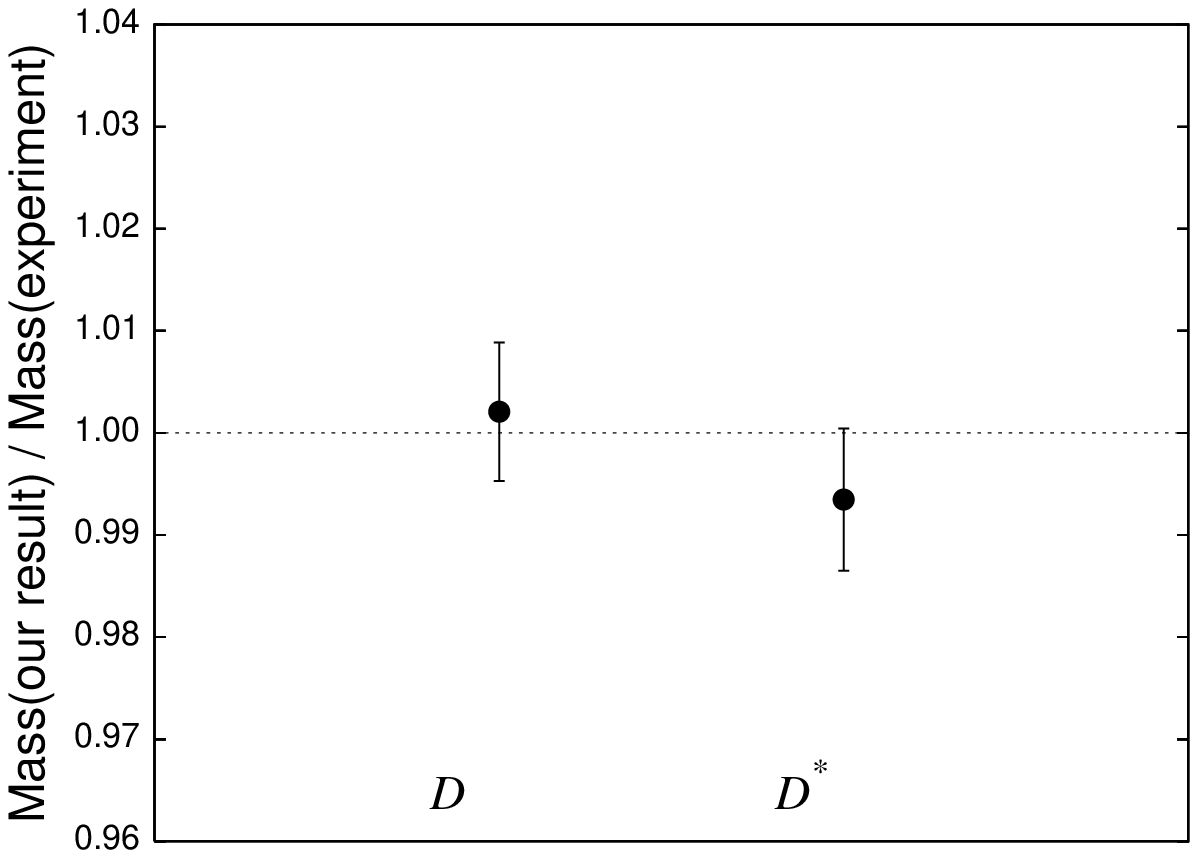}
 \includegraphics[width=75mm]{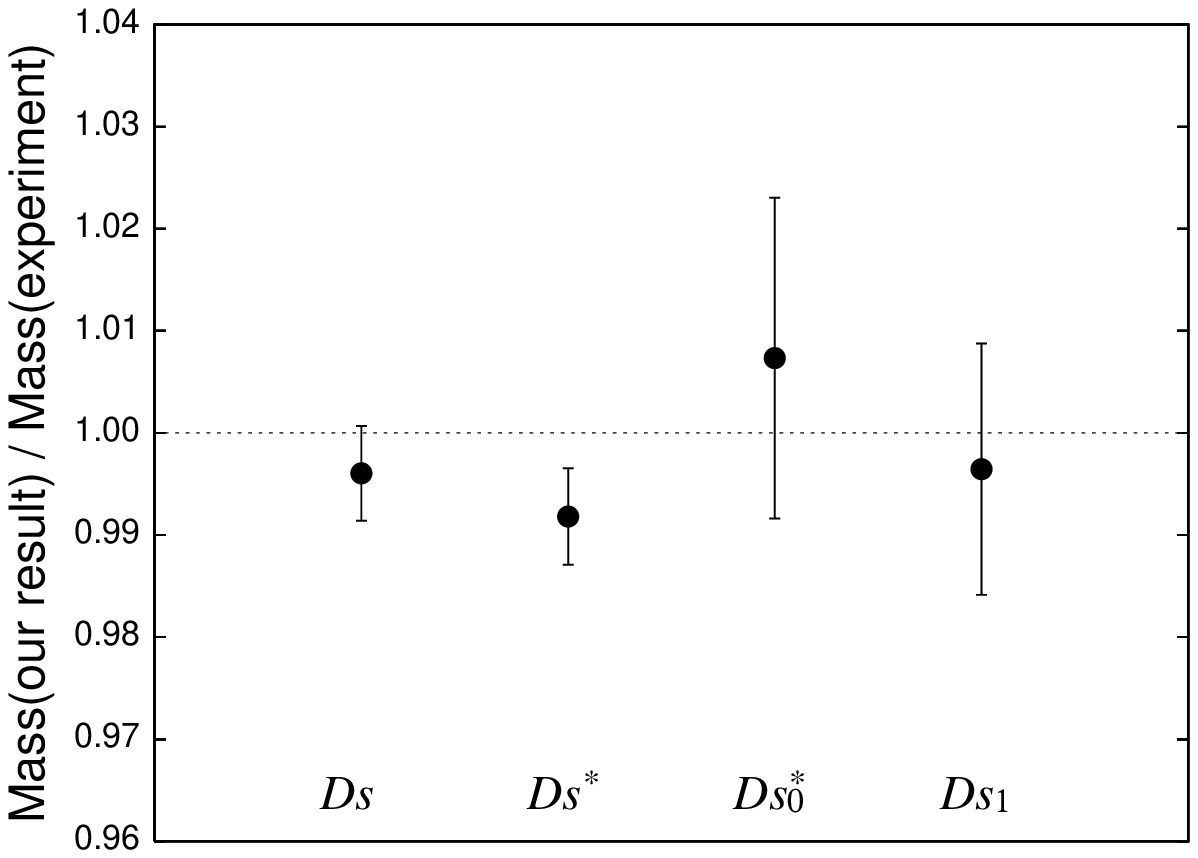}
 \caption{
 Our results for charmed meson masses(left panel)
 and charmed-strange meson masses(right panel)
 normalized by the experimental values.
 }
 \label{figure:mass_ud_charm}
\end{center}
\end{figure}

Our results are summarized in Fig.~\ref{figure:mass_ud_charm} and
in Table~\ref{table:mass_ud_charm} and \ref{table:mass_s_charm}.
All our values for the heavy-light meson quantities are predictions,
because the physical charm quark mass has already been fixed
with the charmonium spectrum.
The experimental spectrum are reproduced in $2 \sigma$ level.
The potential model predicts
the $D_{s0}^*$ meson mass is
above the $D K$ threshold~\cite{potential_model},
which deviates from the experiment significantly.
But, our result does not indicate such a large difference
from the experimental value.
A similar result is obtained in other lattice QCD calculations~\cite{chi_QCD}.
%
%
It should be noticed that our calculation does not cover
$D K$ scattering states yet.
$D K$ contamination for $D_{s0}^*$ and $D_{s1}$ meson masses
can be considerably large.
Further analysis is required to validate
our results for $D_{s0}^*$ and $D_{s1}$ meson spectrum.

\begin{figure}[t]
\begin{center}
 \includegraphics[width=75mm]{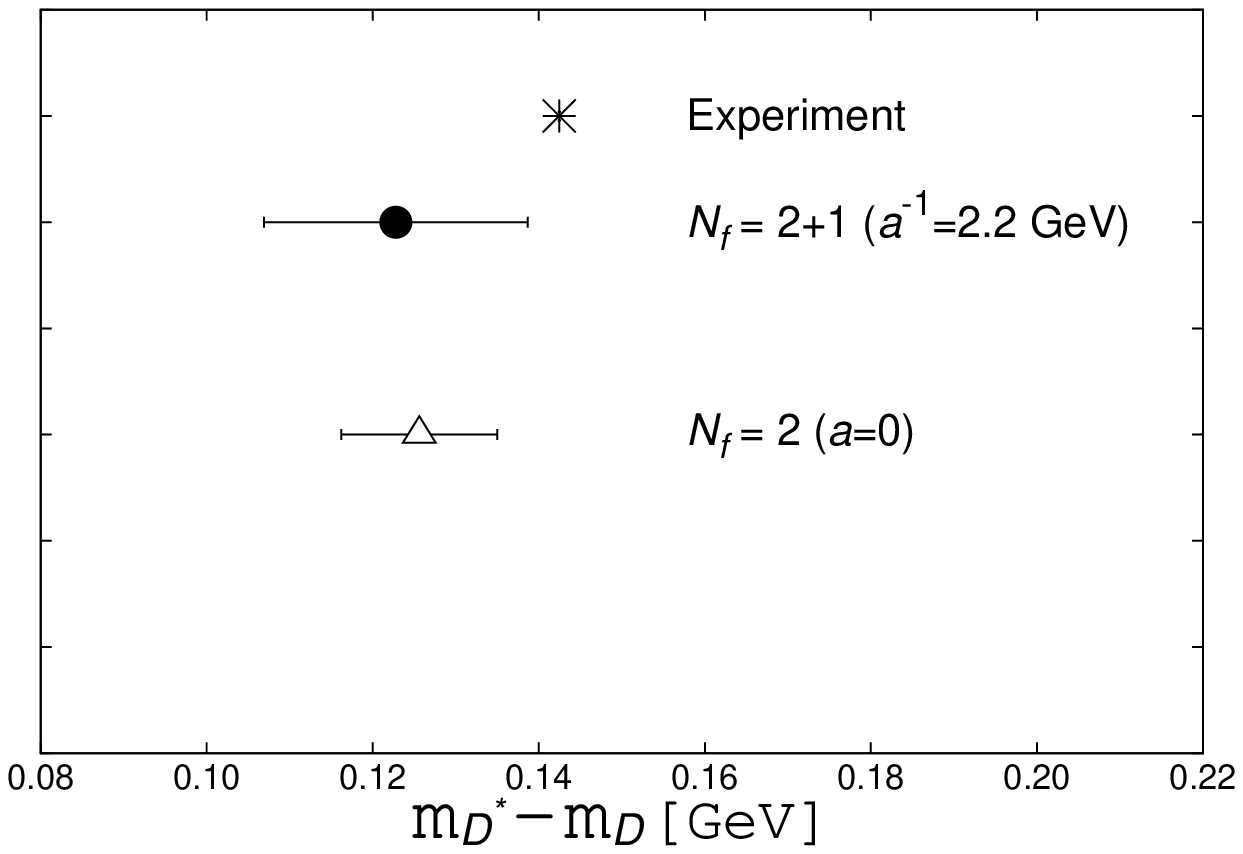}
 \includegraphics[width=75mm]{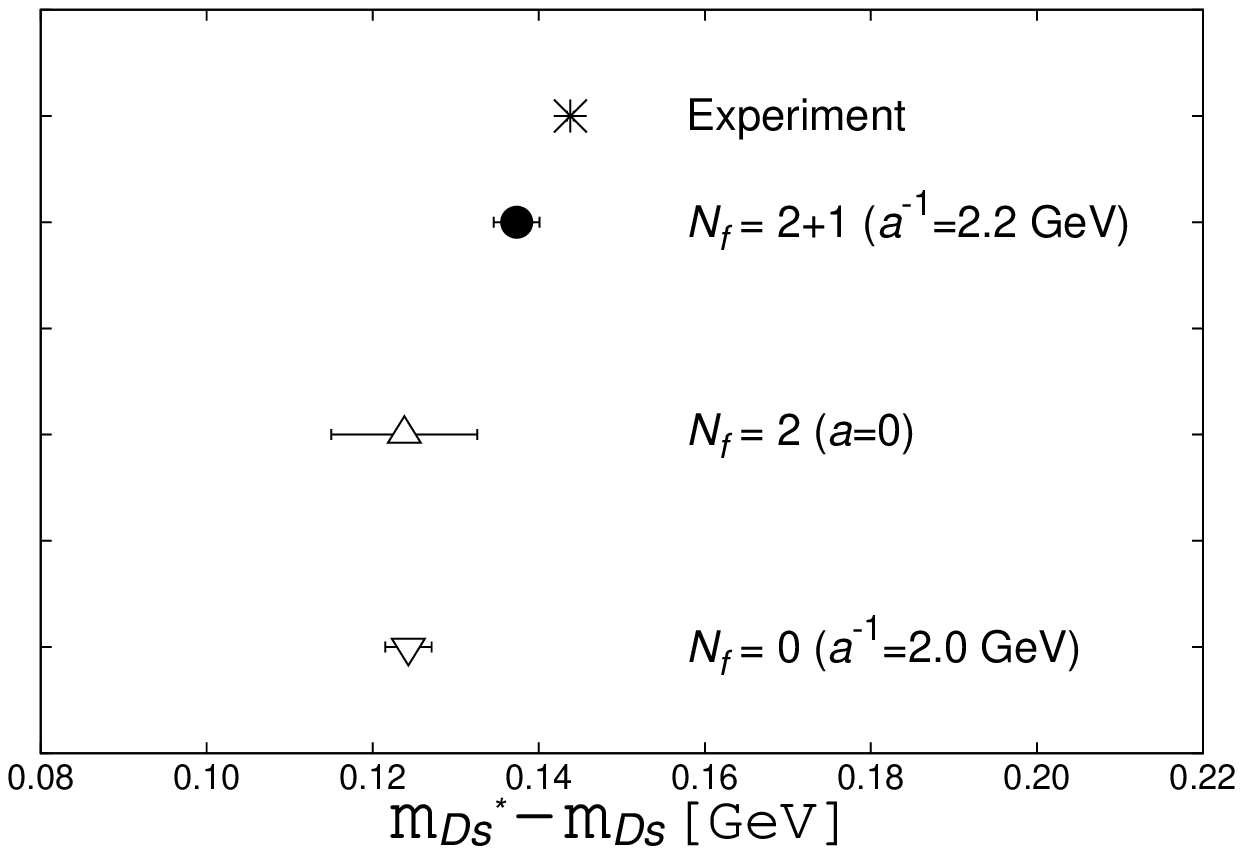}
 \caption{
 Our results for the hyperfine splittings
 of charmed meson(left panel)
 and charmed-strange meson(right panel).
 }
 \label{figure:hyperfine_splitting_ud_s_charm}
\end{center}
\end{figure}

We compare our results for the hyperfine splittings
$m_{D^*}-m_D$ and $m_{D_s}-m_D$ 
with experiments in Fig.~\ref{figure:hyperfine_splitting_ud_s_charm},
where we also plot our previous results
for $N_f=2$ and quenched QCD~\cite{RHQ-N_f_0_2}.
The deviation from the experimental value is
$1.2 \sigma$ for charmed mesons, and
$2.3 \sigma$ for charmed-strange mesons.

\setcounter{equation}{0}
\section{Charmed meson and charmed-strange meson decay constants and CKM matrix elements}
\label{section:result_3}

\begin{figure}[t]
\begin{center}
 \includegraphics[width=75mm]{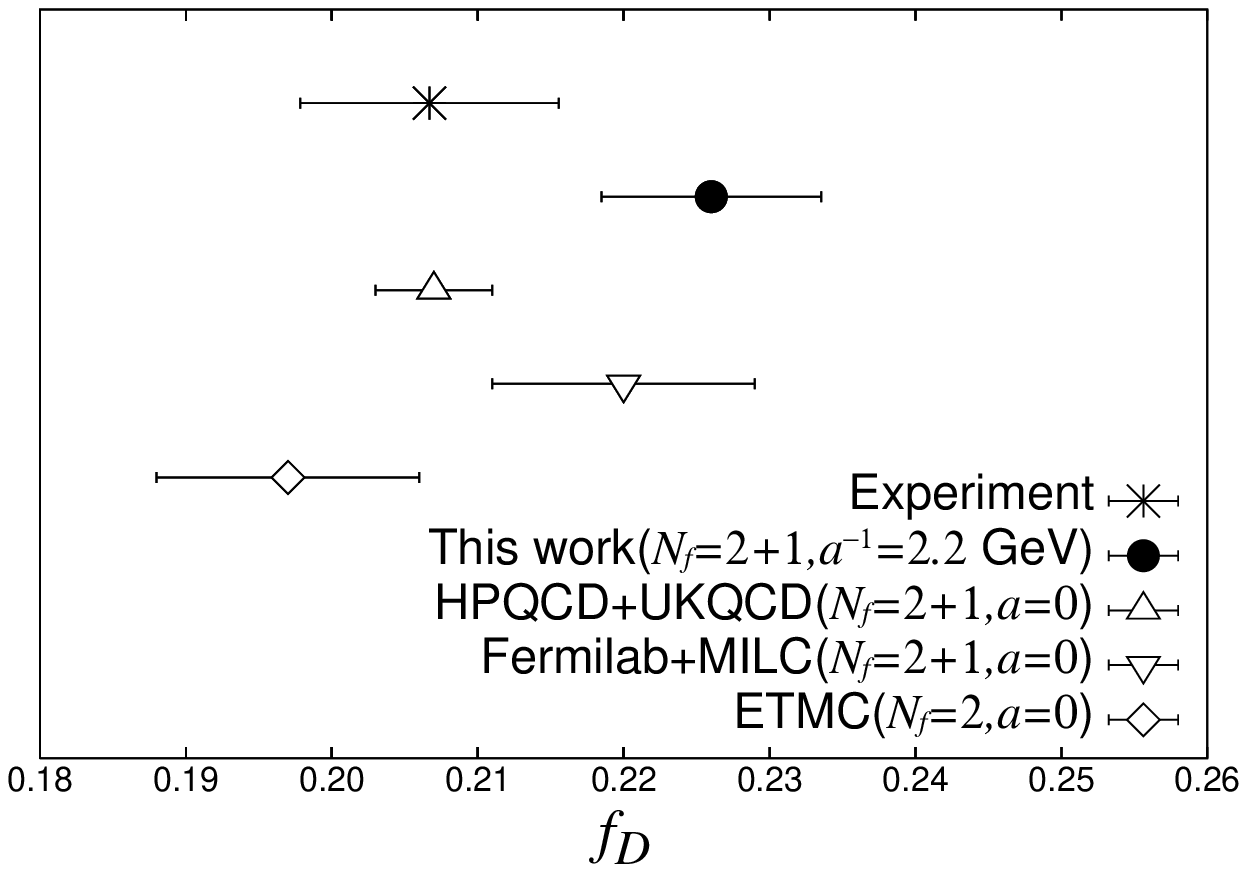}
 \includegraphics[width=75mm]{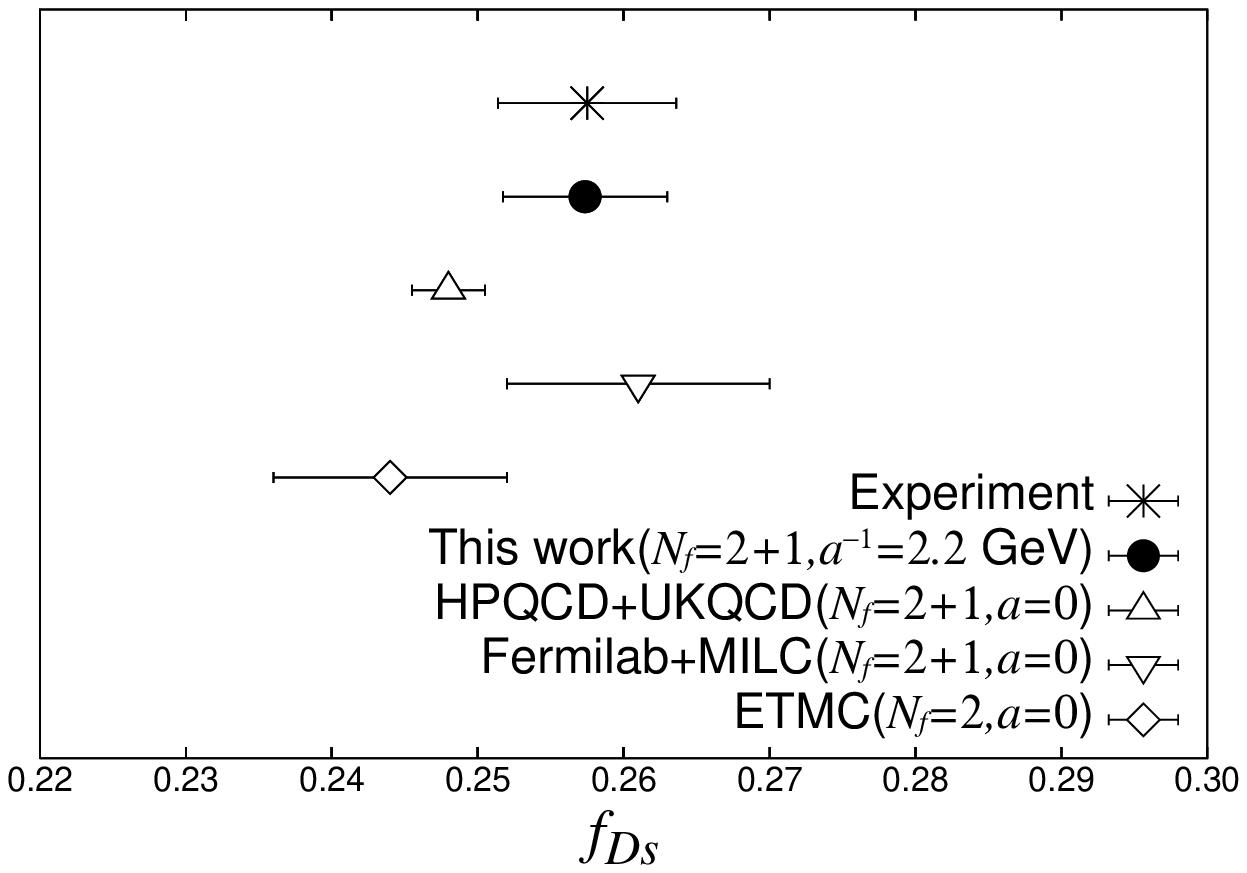}
 \caption{
 Comparison of pseudoscalar decay constants for
 the charmed meson(left panel)
 and charmed-strange meson(right panel).
 }
 \label{figure:f_PS_ud_charm_and_f_PS_s_charm}
\end{center}
\end{figure}

%
%
Table~\ref{table:decay_constants} presents our estimate of
the pseudoscalar decay constants for $D$ and $D_s$ mesons.
Figure~\ref{figure:f_PS_ud_charm_and_f_PS_s_charm}
shows the experimental values~\cite{PDG_2010} and our decay constants,
as well as three recent lattice QCD results: 
HPQCD and UKQCD Collaboration~\cite{HPQCD_1}
using HISQ heavy quark on the MILC 
staggered dynamical configurations,
Fermilab lattice and MILC group~\cite{FNAL_1}
using the Fermilab heavy quark on the MILC configurations, and
ETM Collaboration~\cite{ETMC} who uses the twisted mass formalism.
Our value for $f_{D_s}$ is in accordance
with experiment,
while that for $f_D$ is somewhat larger. 
Comparing four sets of lattice determinations,
we observe, both for $f_D$ and $f_{D_s}$,  
an agreement between our values and those of the Fermilab group,
while there seems to be a discrepancy between our values and
those by the HPQCD and UKQCD Collaboration and ETM Collaboration,
though continuum extrapolation is needed on our part. 
%
%

\begin{figure}[t]
\begin{center}
 \includegraphics[width=75mm]{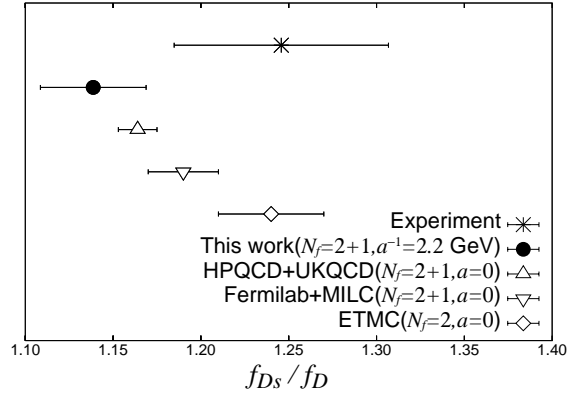}
 \caption{
 Ratios of pseudoscalar decay constants for
 the charmed meson
 and charmed-strange meson.
 }
 \label{figure:f_D_s_over_f_D}
\end{center}
\end{figure}

We plot the ratio of $f_{D_s}$ to $f_D$
in Fig.~\ref{figure:f_D_s_over_f_D}.
Uncertainties coming from the renormalization factors cancel out, 
and that of the lattice cutoff to some extent.
Our result is slightly smaller,
but still $N_f=2+1$ lattice results
are mutually consistent within the errors of a few percent.

\subsection{Estimating the CKM matrix elements}
\label{section:CKM}

The standard model relates $|V_{cd}|$
to the leptonic decay width of the $D$ meson 
$\Gamma(D \rightarrow l \nu)$ by
\begin{eqnarray}
 \Gamma(D \rightarrow l \nu)
 = \frac{G_F^2}{8 \pi} f_{D}^2 m_l^2 m_{D}
   \left( 1 - \frac{m_l^2}{m_{D}^2} \right)^2 |V_{cd}|^2,
\end{eqnarray}
where $G_F$ is the Fermi coupling constant,
and $m_l$ is the lepton mass in the final state.
A lattice determination of the $D$ meson decay constant $f_{D}$
with the experimental value of $\Gamma(D \rightarrow l \nu)$
gives $|V_{cd}|$.
$|V_{cs}|$ can be obtained in the same way.

\begin{figure}[t]
\begin{center}
 \includegraphics[width=75mm]{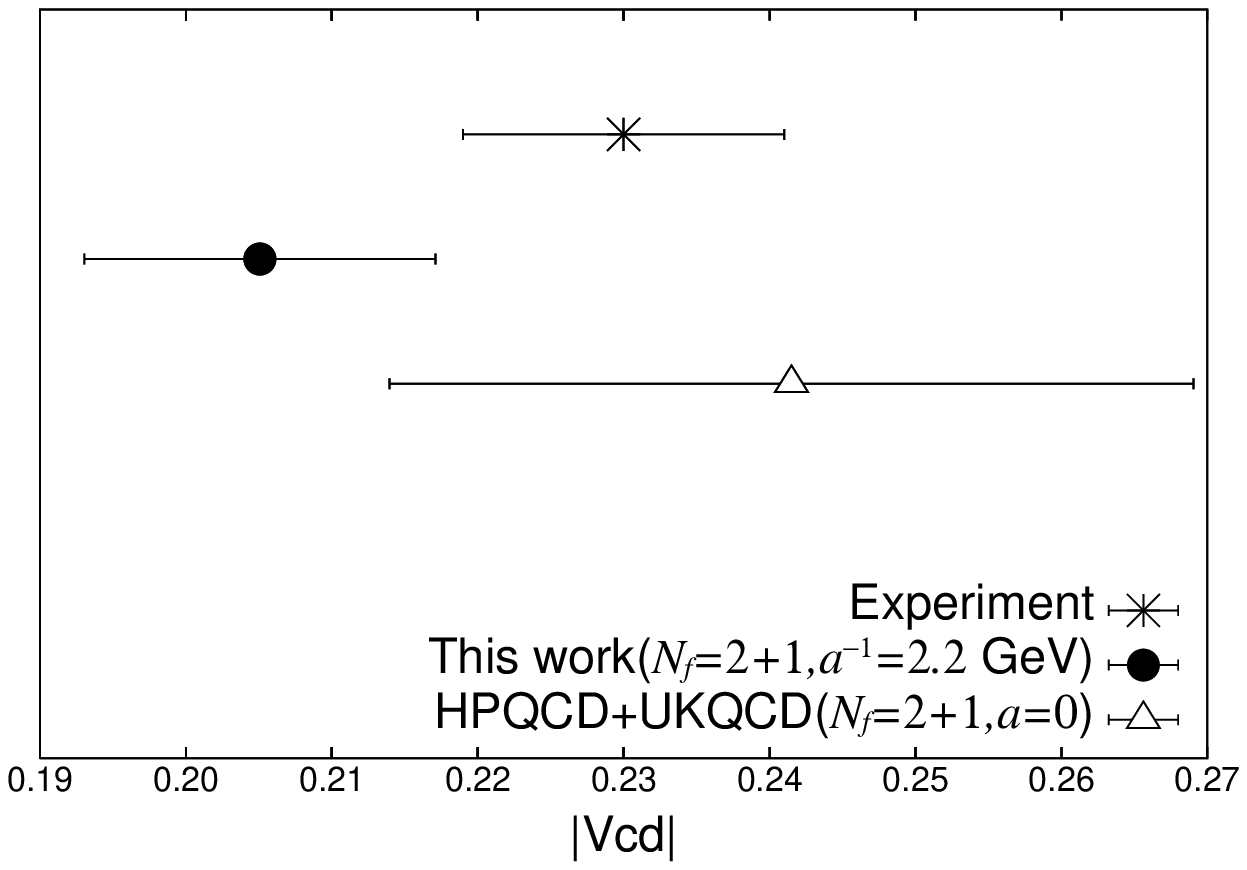}
 \includegraphics[width=75mm]{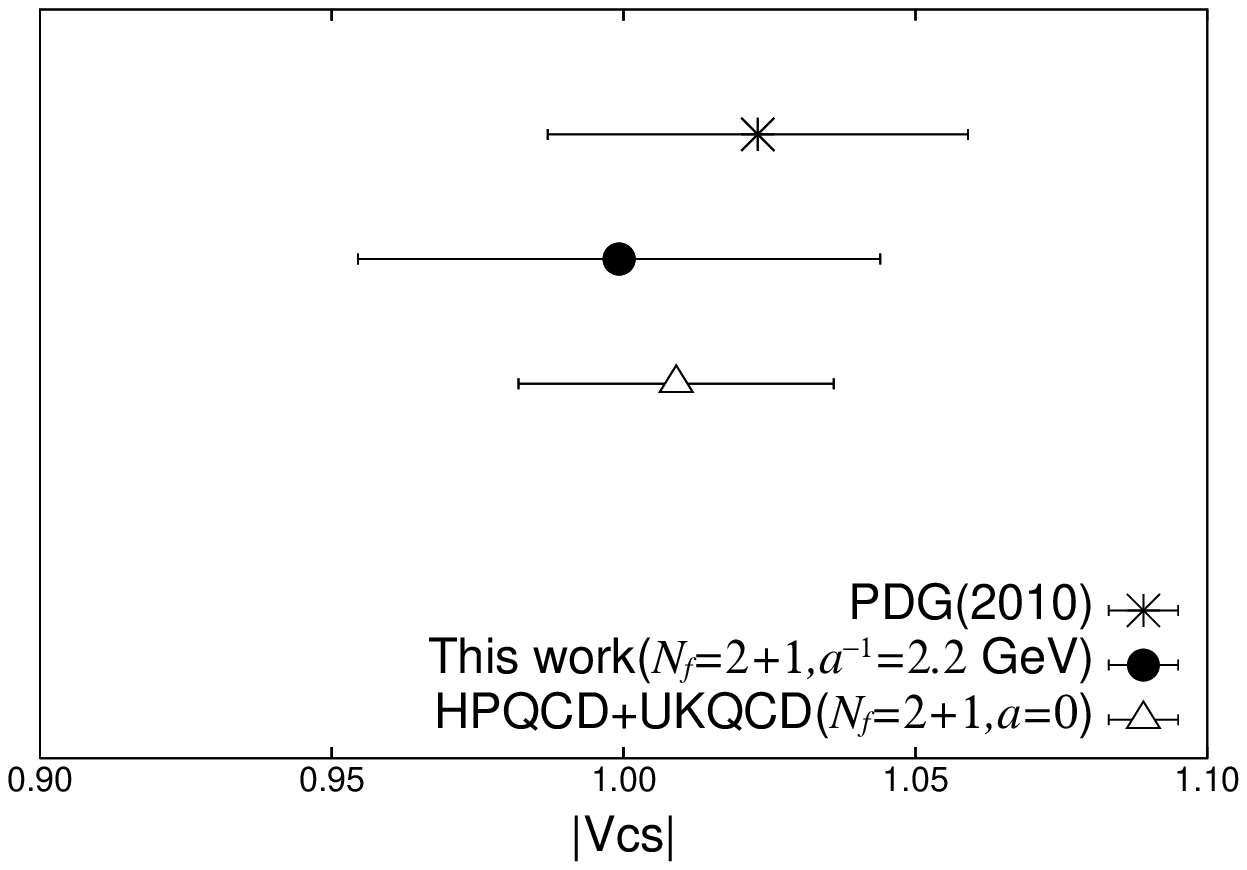}
 \caption{
 Comparison of the CKM matrix elements,
 $|V_{cd}|$(left panel) and $|V_{cs}|$(right panel).
 }
 \label{figure:V_cd_and_V_cs}
\end{center}
\end{figure}

We estimate $|V_{cd}|$ from our $D$ meson mass and decay constant
with the CLEO value of $\Gamma(D \rightarrow l \nu)$~\cite{CLEO_1}.
Up to our heavy quark discretization error of
$O(\alpha_s^2 f(m_Q a)(a \Lambda_{QCD}))$, 
we obtain
\begin{equation}
 |V_{cd}|({\rm lattice}) = 0.205(6)(1)(5)(9),
\end{equation}
where the first error is statistical,
the second is systematic due to the scale determination,
the third is uncertainty of the renormalization factor,
and the forth represents the experimental error of
the leptonic decay width.
For comparison, the PDG value given
by $|V_{cd}|=0.230(11)$ ~\cite{PDG_2010}
is about 10\% larger (see Fig.~\ref{figure:V_cd_and_V_cs}).

Similarly, using the CLEO value of
$\Gamma(D_s \rightarrow l \nu)$~\cite{CLEO_2},
we find 
\begin{equation}
 |V_{cs}|({\rm lattice}) = 1.00(1)(1)(3)(3),
\end{equation}
as compared to $|V_{cs}|= 1.02(4)$ from PDG~\cite{PDG_2010}.

For completeness we also record
the ratio $|V_{cs}| / |V_{cd}|$ for which
the systematic errors are partially dropped out.
\begin{equation}
 \frac{|V_{cs}|}{|V_{cd}|}({\rm lattice})
 = 4.87(14)(0)(0)(27).
\end{equation}
The PDG value is $ |V_{cs}|/|V_{cd}|= 4.45(26)$.

\setcounter{equation}{0}
\section{Conclusion}

We have reported our study of the charm quark system
in $N_f=2+1$ dynamical lattice QCD.
Although carried out at a finite lattice spacing of $a^{-1}=2.194(10)$~GeV,
our results for the spectra of mesons involving charm quarks
are consistent with experiment at a percent level,
and so are those for the decay constants within a few percent accuracy.
These results indicate that the heavy quark mass correction $m_Q a$
in the charm quark system is under control
by the relativistic heavy quark formalism of Ref.~\cite{RHQ_action_Tsukuba}.
Of course, the continuum extrapolation and
further reductions of statistical noises are required
to obtain the result competitive with other approaches in the literature.

From methodological point of view, we have shown that
the realistic heavy quark simulations with the light dynamical quark masses
precisely tuned to the physical values are feasible. 
With the technique of reweighting, configuration generations are needed to be
carried out approximately around the physical point,
and a residual fine tuning to reach the physical point only requires
a much less time consuming evaluation of the quark determinant ratios.  
Combined with the PACS-CS configuration generation
at a smaller lattice spacing of $a^{-1}\approx 3$~GeV underway,
we hope to return to the issue 
of continuum extrapolation for the charm quark system in future.

\section*{Acknowledgments}
Numerical calculations for the present work have been carried out
on the PACS-CS computer
under the ``Interdisciplinary Computational Science Program'' of
Center for Computational Sciences, University of Tsukuba.
This work is supported in part by Grants-in-Aid of the Ministry
of Education, Culture, Sports, Science and Technology-Japan
 (Nos. 18104005, 20105001, 20105002, 20105003, 20105005, 20340047,
 20540248, 21340049, 22105501, 22244018, 22740138).


\clearpage

\begin{table}[t]
\begin{center}
\begin{tabular}{ccccc}
\hline
 $\beta$           &
 $\kappa_{\rm ud}$ & $\kappa_{\rm s}$ &
 \# conf           & MD time
\\ \hline
 1.90              &
 0.13779625        & 0.13663375 &
 80                & 2000
\\ \hline
\end{tabular}
\caption{Simulation parameters.
         MD time is the number of trajectories
         multiplied by the trajectory length.
}
\label{table:statistics}
\end{center}
\end{table}

\begin{table}[t]
\begin{center}
\begin{tabular}{cccccc}
\hline
 $\kappa_{\rm charm}$  & $\nu$     & $r_s$     & $c_B$     & $c_E$
\\ \hline
 0.10959947            & 1.1450511 & 1.1881607 & 1.9849139 & 1.7819512
\\ \hline
\end{tabular}
\caption{Parameters for the relativistic heavy quark action.
}
\label{table:input_parameters_for_RHQ}
\end{center}
\end{table}

\begin{table}[t]
\begin{center}
\begin{tabular}{ccccc}
\hline
                      & $J^{PC}$ & $\Gamma$ operator   & lattice       & experiment
\\ \hline
 $m_{\eta_c}$[GeV]    & $0^{-+}$ & $\gamma_5$          & 2.986(1)(13)  & 2.980(1)
\\ \hline
 $m_{J/\psi}$[GeV]    & $1^{--}$ & $\gamma_i$          & 3.094(1)(14)  & 3.097(0)
\\ \hline
 $m_{\chi_{c0}}$[GeV] & $0^{++}$ & $I$                 & 3.444(33)(15) & 3.415(0)
\\ \hline
 $m_{\chi_{c1}}$[GeV] & $1^{++}$ & $\gamma_i \gamma_5$ & 3.506(30)(15) & 3.511(0)
\\ \hline
 $m_{h_{c}}$[GeV]     & $1^{+-}$ & $\gamma_i \gamma_j$ & 3.510(42)(15) & 3.525(0)
\\ \hline
\end{tabular}
\caption{
 Charmonium spectrum in GeV units.
 The first error is statistical,
 and the second is systematic
 from the scale determination.
 Experimental data are also listed~\cite{PDG_2010}.
}
\label{table:mass_charmonium}
\end{center}
\end{table}

\begin{table}[t]
\begin{center}
\begin{tabular}{ccccc}
\hline
                & $J^P$  & $\Gamma$ operator & lattice      & experiment
\\ \hline
 $m_{D}$[GeV]   & $0^-$  & $\gamma_5$        & 1.871(10)(8) & 1.865(0)
\\ \hline
 $m_{D^*}$[GeV] & $1^-$  & $\gamma_i$        & 1.994(11)(9) & 2.007(0)
\\ \hline
\end{tabular}
\caption{
 Charmed meson
 mass spectrum in GeV units.
 The first error is statistical,
 and the second is systematic
 from the scale determination.
 Experimental data are also listed~\cite{PDG_2010}.
}
\label{table:mass_ud_charm}
\end{center}
\end{table}

\begin{table}[t]
\begin{center}
\begin{tabular}{ccccc}
\hline
                     & $J^P$ & $\Gamma$ operator   & lattice       & experiment
\\ \hline
 $m_{D_s}$[GeV]      & $0^-$ & $\gamma_5$          & 1.958(2)(9)   & 1.968(0)
\\ \hline
 $m_{D_s^*}$[GeV]    & $1^-$ & $\gamma_i$          & 2.095(3)(10)  & 2.112(1)
\\ \hline
 $m_{D_{s0}^*}$[GeV] & $0^+$ & $I$                 & 2.335(35)(10) & 2.318(1)
\\ \hline
 $m_{D_{s1}}$[GeV]   & $1^+$ & $\gamma_i \gamma_5$ & 2.451(28)(11) & 2.460(1)
\\ \hline
\end{tabular}
\caption{
 Charmed-strange meson
 mass spectrum in GeV units.
 The first error is statistical,
 and the second is systematic
 from the scale determination.
 Experimental data are also listed~\cite{PDG_2010}.
}
\label{table:mass_s_charm}
\end{center}
\end{table}

\begin{table}[t]
\begin{center}
\begin{tabular}{ccc}
\hline
                    & lattice        & experiment
\\ \hline
 $f_{D}$[MeV]       & 226(6)(1)(5)   & 206.7(8.9)
\\ \hline
 $f_{D_s}$[MeV]     & 257(2)(1)(5)   & 257.5(6.1)
\\ \hline
 $f_{D_s}/f_{D}$    & 1.14(3)(0)(0)  & 1.25(6)
\\ \hline
\end{tabular}
\caption{
 Our results for decay constants of
 $D$ meson and $D_s$ meson.
 The first error is statistical,
 the second is systematic
 from the scale determination,
 and the third is from the renormalization factor.
 Experimental data are also listed~\cite{PDG_2010}.
}
\label{table:decay_constants}
\end{center}
\end{table}

\clearpage

\begin{figure}[t]
\begin{center}
 \includegraphics[width=75mm]{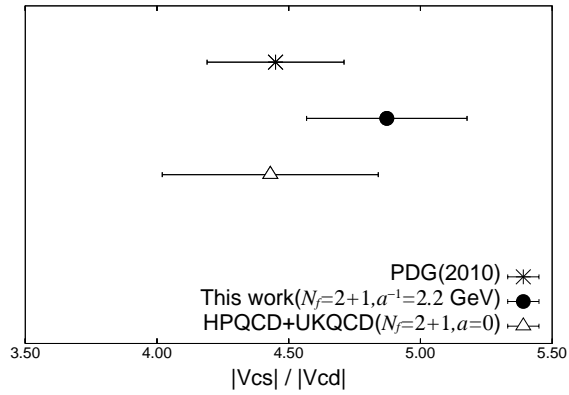}
 \caption{
 Ratio of the CKM matrix elements,
 $|V_{cs}|$ and $|V_{cd}|$.
 }
 \label{figure:V_cd_over_V_cs}
\end{center}
\end{figure}


\end{document}